\documentclass[preprint,12pt]{elsarticle}

\usepackage{graphicx}
\usepackage[all]{xy}
\usepackage{amsmath}
\usepackage{amssymb}
\usepackage{epstopdf}
\usepackage[colorlinks=true, pdfstartview=FitV, linkcolor=blue, citecolor=red, urlcolor=magenta]{hyperref}
\usepackage{latexsym}
\usepackage{amsfonts}
\usepackage{subeqnarray}

\newcommand{\be}{\begin{equation}}
\newcommand{\ee}{\end{equation}}
\newcommand{\ben}{\begin{eqnarray}}
\newcommand{\een}{\end{eqnarray}}
\newcommand{\bsen}{\begin{subeqnarray}}
\newcommand{\esen}{\end{subeqnarray}}

\newcommand{\bb}{\bibitem}

\newcommand{\bea}{\begin{eqnarray}}
\newcommand{\eea}{\end{eqnarray}}

\newcommand{\rag}{\rangle}

\journal{Physica A}

\begin{document}

\begin{frontmatter}



\title{Harmonic oscillators from displacement operators and thermodynamics}


\author[addr1,addr2]{F. A. Brito}
\ead{fabrito@df.ufcg.edu.br}
\author[addr2]{F. F. Santos }
\ead{fabiano.ffs23@gmail.com}
\author[addr1]{J. R. L. Santos}
\ead{joaorafael@df.ufcg.edu.br}

\address[addr1]{Departamento de F\'\i sica, Universidade Federal de Campina Grande, Caixa Postal 10071,
58109-970  Campina Grande, Para\'\i ba, Brazil}
\address[addr2]{Departamento de F\'\i sica, Universidade Federal da Para\'iba, Caixa Postal 5008, 58051-970, Jo\~ ao Pessoa, Para\' iba, Brazil.}

\begin{abstract}
In this investigation, the displacement operator is revisited. We established a connection between the Hermitian version of this operator with the well-known Weyl ordering. Besides, we characterized the quantum properties of a simple displaced harmonic oscillator, as well as of a displaced anisotropic two-dimensional non-Hermitian harmonic oscillator. By constructing the partition functions for both harmonic oscillators, we were able to derive several thermodynamic quantities from their energy spectra. The features of these quantities  were depicted and analyzed in details. 

\end{abstract}


\end{frontmatter}




\section{Introduction}
\label{sec_1}

A proper way to quantize the space-time has been a major challenge in physics since the early days of non-relativistic quantum mechanics. One beautiful approach which quantizes the space-time in relativistic quantum field theory, was proposed by Snyder in his seminal paper  \cite{Hs}. There he points that Lorentz invariance imposes no restriction to the quantization of the space-time itself, and also that a natural way to develop such a quantization is through the definition of a noncommutative space operator. One example of this class of space operator is the one introduced by Costa Filho {\it et al.} \cite{r1}, which was denominated displacement operator. In the mentioned work, the authors shown how the displacement operator can describe the quantum dynamics of a position dependent mass particle subject to a null and to a constant potential. Such a description appears as consequence of the infinitesimal displacement operator be described as an effective position dependent linear momentum, which then is equivalent to a position dependent mass for a quantum particle.

A Hermitian version of the displacement operator was proposed in  \cite{r2} where the energy states and  observables for the dynamics of the effective quantum particle were also determined. An interesting question to address when a quantum operator is established is the ordering ambiguity in the product $\hat{x}\,\hat{p}$. In this study we propose a new route to characterize the operator introduced in \cite{r2}, based on a generalized version of the  Weyl ordering \cite{r9}. Besides, we show how the  displacement operator from \cite{r2} is a direct consequence of a specific ordering.  In our discussions, we also revisit the calculation of the eigenvalues and of the eigenstates for a displaced harmonic oscillator. 

The generalized version of the displacement operator enable us to discuss non-hermitian quantum systems, which also break the $PT$ symmetry (parity and time reversal). The parity, and time reversal operations can be summarized as $P:\,\, x\rightarrow -x\,,\,\, y\rightarrow -y$, and $T:\,\,t\rightarrow-t\,,\,\,i\rightarrow-i$, respectively. In their seminal work, Bender and Boettcher \cite{bender} proposed the invariance under $PT$ symmetry as a criteria to generate real spectrum. 

 The main motivation to deal with $PT$ symmetric non-Hermitian quantum mechanics is to derive several new Hamiltonians which can describe physical system, besides, the $PT$ symmetry can be interpreted as a condition stronger than hermiticity to obtain energies spectra to quantum systems \cite{bender_2003}. It is relevant to point that $PT$ symmetric non-Hermitian operators were used to describe oscillators with time dependent mass and frequency \cite{dutra_2005}. Such oscillators are commonly applied to quantum optics, quantum chemistry and to describe the electromagnetic field inside of a Fabry-P\'erot cavity \cite{dutra_2005,colegrave_1983}. Moreover, $PT$ symmetric systems was also investigated in the context of classical field theory, where topological defects with real energies were found \cite{dutra_2007}.

In contrast with the studies of Bender and Boettcher, in our work we deal with a two-dimensional displaced version of the problem presented in \cite{bender} which also breaks the $PT$ symmetry. As it is shown, despite the fact that the $PT$ symmetry is not obeyed, we can find a real spectrum condition which enable us to determine real energy values for this oscillator. 

In our investigation besides the determination of this real spectrum condition, we also find the eigenvalues as well as the eigenstates for this two-dimensional harmonic oscillator. In order to complete the characterization of our spectra, we study the thermodynamic properties of both the simple displaced and the two-dimensional non-Hermitian displaced harmonic oscillators. The procedure adopted to establish partition functions for both oscillators follows the same lines of Morse oscillators applied to molecules \cite{strekalov}. The thermodynamic quantities from these partition functions unveil some remarkable features about these oscillators, such as internal energies coinciding with a two-level system, and specific heats with Schottky anomaly. Moreover, some free constants of the two-dimensional non-Hermitian oscillator can be used as a fine-tunning for the thermodynamic quantities.

Our discussions in this article are divided in the sections: In section \ref{sec_2} we work with the general aspects about the displacement operator, and we also analyze its Hermitian properties via the generalized Weyl ordering. Section \ref{sec_3} is dedicated to revisit the computation of the eigenstates and of the eigenvalues for the simple displaced harmonic oscillator. Besides, in section \ref{sec_4} we study the quantum mechanical features of the displaced anisotropic two-dimensional non-Hermitian harmonic oscillator. Section \ref{sec_5} is dedicated to the statistical mechanics of both oscillators, where we describe carefully the procedures to derive their partition functions and we plot the thermodynamic quantities. Finally, in section \ref{sec_6} we present our conclusions and perspectives.

\section{On the displacement operator}
\label{sec_2}

This section is focused in a brief review about the definition of the displacement operator, which was introduced by \cite{r1}, and discussed in more details in \cite{r2}. Let us work with a well localized state around the position $x$, which can be displaced for a new one characterized by $x+a+\gamma\,a\,x$. It is relevant to say that despite the displacement of the state, its physical properties are fixed. Moreover, if $\gamma$ parameter is null, we are talking about an usual translation, on the other hand, the case $\gamma \neq 0$ represents a position dependent displacement.

These previous statements allow one to establish that a displacement operator, here denoted as  $T_{\,\gamma} (a)$, obeys the following property
\be \label{eq1.1}
T_{\,\gamma}(a)\, | x \rag = |\,x+a+\gamma\,a\,x \rag \,,
\ee
when it is applied to a state $| x \rag$. So, considering two infinitesimal displacements $dx^{\,\prime}$, and $dx^{\,\prime\prime}$ we directly determine that
\be \label{eq1.2}
T_\gamma(dx^{\,\prime})\,T_\gamma(dx^{\,\prime\prime})=T_\gamma(dx^{\,\prime}+dx^{\,\prime\prime}+\gamma\,dx^{\,\prime}\,dx^{\,\prime\prime})\,,
\ee
unveiling a non-additive kind of displacement. 

The behavior presented by $(\ref{eq1.1})$ is similar to the product involving two q-exponential functions $\exp_q(y)$ \cite{r3}, which is explicitly written as
\be \label{eq1.3}
\exp_q(a)\,\exp_q(b)=\exp_q\,\left[a+b+(1-q)\,a\,b\right]\,.
\ee
Therefore, the displacement operator $T_{\,\gamma}(a)$ with $\gamma=1-q$, can be thought as the generator of the q-exponentials. Such special exponential functions are the essential ingredient of the so-called Tsallis non-extensive thermodynamics \cite{r5}, and they can be applied in different physical scenarios, as those investigated in \cite{r6,r7,r8}.

Furthermore, the inverse displacement operator can be defined as
\be \label{eq1.4}
T_{\,\gamma}^{\,-1}(dx)\,|x \rag = \left|\frac{x-dx}{1+\gamma\,dx}\right \rag\,,
\ee
and we can observe that if there is no translation, or in other works, if we take the limit
\be \label{eq1.5}
\lim_{dx\rightarrow 0}\,T_{\,\gamma}(dx)= 1\,,
\ee
the displacement operator becomes an identity matrix.

One way to characterize such a displacement consists in the definition
\be \label{eq1.6}
T_{\,\gamma}(dx)=1-\frac{i}{\hbar}\,\hat{p}_{\,\gamma}\,dx\,,
\ee
moreover, from Eq. $(\ref{eq1.1})$ we see that
\ben\label{eq1.7}
\hat{x}T_{\,\gamma}(dx)\,|x\rag &=& \hat{x}\,|x+dx+\gamma\,x\,dx \rag\,, \\ \nonumber
&=&
\left(x+dx+\gamma\,x\,dx\right)\,|x+dx+\gamma\,x\,dx \rag \,,
\een
\be \label{eq1.8}
T_{\,\gamma}(dx)\hat{x}\,|x\rag = x\,|x+dx+\gamma\,x\,dx \rag\,,
\ee
therefore
\ben \label{eq1.9}
\left[\hat{x},T_{\,\gamma}(dx)\right]\,|x\rag &=& dx\,(1+\gamma\,x)\,|x+dx+\gamma\,x\,dx \rag\,,\\ \nonumber
&\approx &
dx\,(1+\gamma\,x)\,|x\rag\,,
\een
up to first-order in $\gamma$. By combining the last result with $(\ref{eq1.6})$, we find that the commutation relation involving the momentum and the $\hat{x}$ operator is
\be \label{eq1.10}
\left[\hat{x},\hat{p}_{\,\gamma}\right]=i\,\hbar\,(1+\gamma\,x)\,,
\ee
for this specific algebra.

Inspired by this commutation relation, the displacement momentum operator $\hat{p}_{\,\gamma}$ was defined by \cite{r1,r2}, as
\be \label{eq1.11}
\hat{p}_{\gamma}=\frac{\hbar}{i}\left(1+\gamma\,x\right)\,\frac{d}{d\,x}\,,
\ee
and
\be \label{eq1.12}
\hat{p}_{\,\gamma}=\frac{\hbar}{i}\left(\left(1+\gamma\,x\right)\,\frac{d}{d\,x}+\frac{\gamma}{2}\right)\,,
\ee
respectively. By following the procedures adopted in \cite{r2}, it is possible to show that $\hat{p}_{\gamma}$ from $(\ref{eq1.12})$ is Hermitian. Such a property yields to the unitary of $T_{\gamma}$ operation, or in another words, $T_{\,\gamma}^{\dagger}(dx)\,T_{\,\gamma}(dx) = 1$.

As it is known the standard quantum mechanics presents the so-called problem of the ordering ambiguity in the product $\hat{x}\,\hat{p}$. A very interesting debate about this subject involving names like Born, Dirac, Jordan, von-Newmann,  and Weyl  can be found in \cite{r9}. Moreover, this ambiguity commonly appears in quantum mechanical systems with mass dependent potentials, as one can see in  \cite{r10}, as well as, in quantum semiconductor heterostructures  \cite{r11}. 

Guided by the problem of the ordering ambiguity, and inspired by the simplest generalization for the Weyl ordering presented in \cite{r9}, we introduce the following $\hat{p}_{\,\gamma}$ operator
\be \label{eq1.13}
\hat{p}_{\,\gamma}=\frac{\hbar}{i}\left[\left(\xi+\gamma\,\bar{\alpha}\,x\right)\,\frac{d}{d\,x}+\gamma\,\bar{\beta}\,\frac{d}{d\,x}\,x\right]\,; \qquad \bar{\alpha}+\bar{\beta}=1\,.
\ee
Let us check the proper conditions to find a Hermitian version for this operator. In order to proceed, we compute the inner product
\ben \label{eq1.14}
\langle f| \hat{p}_{\gamma} |g \rangle &=& \int\,dx\,f^{\,*}\left(\frac{\hbar}{i}\left[\left(\xi+\gamma\,\bar{\alpha}\,x\right)\,\frac{d\,g}{d\,x}+\gamma\,\bar{\beta}\,\frac{d}{d\,x}\,(x\,g)\right]\right)\,; \\ \nonumber
&=& \int\,dx\,f^{\,*}\left(\frac{\hbar}{i}\left[\left(\xi+\gamma\,x\right)\,\frac{d\,g}{d\,x}+\gamma\,\bar{\beta}\,g\right]\right)\,,
\een
where we use the fact that $\bar{\alpha}+\bar{\beta}=1$. So, integrating the last equation  by parts we find
\be \label{eq1.15}
\langle f| \hat{p}_{\gamma} |g \rangle = \int\,dx\,\left(-\frac{\hbar}{i}\right)\left[(\xi+\gamma\,x)\,\frac{d\,f^{\,*}}{d\,x}+\gamma\,(1-\bar{\beta})f^{\,*}\right]\,g+\mbox{boundary terms}\,.
\ee
From this last expression, we can observe that $\hat{p}_{\gamma}$ is Hermitian, if $\xi$ is real and if $\bar{\alpha}=\bar{\beta}=1/2$. Moreover, we set the boundary terms equal to zero. Therefore, the previous parameters enable us to rewrite $(\ref{eq1.15})$ as
\ben \label{eq1.16}
\langle f| \hat{p}_{\gamma} |g \rangle &=& \int\,dx\,\left(-\frac{\hbar}{i}\right)\left[(\xi+\gamma\,x)\,\frac{d\,f^{\,*}}{d\,x}+\frac{\gamma}{2}\,f^{\,*}\right]\,g \\ \nonumber
&=& \int\,dx\left(\hat{p}_\gamma\,f\right)^{\,\dagger}\,g \\ \nonumber
&=& \langle \hat{p}_{\gamma}\,f|g\rangle\,,
\een
proving that in this case $\hat{p}_{\gamma}$ is in fact Hermitian. 

This new representation of $\hat{p}_{\,\gamma}$ (Eq. $(\ref{eq1.13})$), is consistent with $(\ref{eq1.10})$, as well as, with Eq.  $(\ref{eq1.12})$. The mentioned values for $\bar{\alpha}$, and $\bar{\beta}$ are denominated Weyl order, representing a symmetric ordering between $\hat{x}$, and $\hat{p}$. 

Therefore, the constant $-i\,\hbar\,\gamma/2$ added in $\hat{p}_{\,\gamma}$ (Eq. $(\ref{eq1.12})$),  can be viewed as a consequence of the  Weyl order, which is also responsible for the Hermitian condition of  $T_{\,\gamma}$. There are several physical motivations to use this special order, in particular, we can point its broadly application in the path-integrals formalism to determine the medium point prescription \cite{r12,r13}. Furthermore,  despite the fact that different values of $\bar{\alpha}$, and $\bar{\beta}$ lead to non-Hermitian displacement operators, we are able to find some physical interesting results, as those pointed by \cite{r1}.

\section{The displaced harmonic oscillator}
\label{sec_3}

In this section we review some generalities about the displaced simple harmonic oscillator, following the procedures adopted by \cite{r1,r2, Raimundo}. We start our investigation establishing that,
\be \label{eq2.1}
\hat{p}_{\,\gamma}=\frac{\hbar}{i}\,{\cal D}_{\,\gamma}\,,
\ee
with
\be \label{eq2.2}
{\cal D}_{\gamma}=\left(1+\gamma\,x\right)\,\frac{d}{d\,x}+\frac{\gamma}{2}\,,
\ee
as the Hermitian displacement momentum operator. 
Therefore, 
\be \label{eq2.4}
\hat{H}=\frac{\hat{p}_{\,\gamma}^{\,2}}{2\,m}+V(x)\,,
\ee
is the Hamiltonian operator for momentum $\hat{p}_{\,\gamma}$. Such an operator results in the time-independent Schroedinger equation
\be \label{eq2.5}
-\frac{\hbar^2}{2m}\,{\cal D}_{\,\gamma}^{\,2}\,\phi(x)+V(x)\,\phi(x)=E\,\phi(x)\,,
\ee
where $\phi(x)$ is the wave-function. As it is known, the harmonic oscillator is described by the potential
\be \label{eq2.6}
V(x)=\frac{m\omega^{2}x^{2}}{2}\,,
\ee
then, by applying the change of variables $u=1+\gamma\,x$, we directly determine ${\cal D}_{\,\gamma}$ as
\be \label{eq2.7}
{\cal D}_{\,\gamma}=u\gamma\,\frac{d}{d\,u}+\frac{\gamma}{2}\,;\qquad {\cal D}_{\,\gamma}^{\,2}=u^2\gamma^2\,\frac{d^{\,2}}{d\,u^2}+2u\gamma^2\,\frac{d}{d\,u}+\frac{\gamma^2}{4}\,,
\ee
Thus, the previous ingredients results in the following form for $(\ref{eq2.5})$:
\be \label{eq2.8}
-\frac{\hbar^{2}}{2m}\,\left[u^2\gamma^2\,\phi_{\,u\,u}+2u\gamma^2\,\phi_{\,u}+\frac{\gamma^2}{4}\,\phi\right]+\frac{m\,\omega^2}{2\,\gamma^2}(u-1)^2\,\phi=E\,\phi\,,
\ee
which can be rewritten as
\be \label{eq2.9}
u^2\,\phi_{\,u\,u}+2u\,\phi_{\,u}-\,\widetilde{\omega}^2(u-1)^2\,\phi=\widetilde{E}\,\phi\,,
\ee
where
\be \label{eq2.10}
\widetilde{E}=-\frac{2\,m}{\hbar^2\gamma^2}\,E-\frac{1}{4}\,;\qquad \widetilde{\omega}^2=\frac{m^2\,\omega^2}{\hbar^2\,\gamma^4}\,.
\ee

Considering a new change of variables, given by $u=z+1$, $(\ref{eq2.10})$ is represented as
\be \label{eq2.11}
(z+1)^2\,\phi_{\,z\,z}+2(z+1)\,\phi_{\,z}-\widetilde{\omega}^{\,2}\,z^2\,\phi=\widetilde{E}\,\phi\,,
\ee
which can be simplified if we use $\phi(z)\rightarrow \phi/(z+1)$, yielding to
\be \label{eq2.12}
(z+1)^2\,\phi_{\,z\,z}-\,\widetilde{\omega}^{\,2}\,z^2\,\phi=\widetilde{E}\,\phi\,,
\ee
The analytical solution of the last equation is given in terms of the associated Laguerre polynomials, and its explicit form is
\be \label{eq2.13}
\phi=e^{-\widetilde{\omega}(z+1)}(z+1)^{\widetilde{\omega}-n}L_{n}^{\kappa}[2\widetilde{\omega}(z+1)]\,,
\ee
where $n$, and $\kappa$ are
\be \label{eq2.14}
n=-\frac{1}{2} +\widetilde{\omega}-\frac{1}{2}\sqrt{1+4\widetilde{E}+4\widetilde{\omega}^{2}}\,;\qquad \kappa=\sqrt{1+4\widetilde{E}+4\widetilde{\omega}^{2}}\,,
\ee
moreover, $n$ must be an integer. The associated Laguerre polynomials are also presented  the context of nuclear shell models where the nucleons are subject to an interaction mediated by a harmonic oscillator, see \cite{dasferbel} for more details. Besides,  in terms of the original variables,  Eq. $(\ref{eq2.13})$ is such that
\be
\phi(x)=e^{-\frac{m\omega}{\hbar\,\gamma^2}\,(1+\gamma\,x)}\,(1+\gamma\,x)^{-\frac{m\omega}{\hbar\,\gamma^2}-n-1}\,L_{n}^{\kappa}\,\left[2\,\frac{m\omega}{\hbar\gamma^2}\,(1+\gamma\,x)\right]\,.
\ee

Following the procedures adopted by \cite{dasferbel} in $(\ref{eq2.14})$, we determine
\be \label{eq2.15}
\widetilde{E}=n(n+1)-\widetilde{\omega}(1+2n);\,\,\,\,E=\frac{\hbar^{2}\gamma^{2}}{m}\widetilde{\omega}\left(n+\frac{1}{2}\right)-\frac{\hbar^{2}\gamma^{2}}{2\,m}\left(n\left(n+1\right)+\frac{1}{4}\right)\,,
\ee
as the spectrum for the displaced harmonic oscillator, whose form given by the  original variables is
\be \label{eq2.16}
E=\hbar\omega\left(n+\frac{1}{2}\right)-\frac{\hbar^{2}\gamma^{2}}{2\,m}\left(n+\frac{1}{2}\right)^{2}\,.
\ee

The previous spectrum was derived by Costa Filho {\it et al.} \cite{Raimundo}, where the authors traced a parallel between the Schroedinger equation for the displaced harmonic oscillator, and the Morse potential.

\section{The displaced anisotropic harmonic oscillator}
\label{sec_4}

In this section we are going to approach a displaced anisotropic two-dimensional non-Hermitian harmonic oscillator. Such a system also breaks the so-called  $PT$ symmetry (parity and time reversal).  We show below how this non-Hermitian oscillator subject to a constraint has real spectrum, despite it breaks the $PT$ symmetry. 

Moving further, the potential for the anisotropic harmonic oscillator can be defined as
\be \label{eq3.1}
V(x,y)=\frac{1}{2}\,m\,\omega_1^2\,x^2+\frac{1}{2}\,m\,\omega_2^2\,y^2+(\lambda_1+i\,\sigma_1)\,x+(\lambda_2+i\,\sigma_2)\,y\,,
\ee
where $\lambda_j$, and $\sigma_j$ ($j=1,2$) are real constants, and the $PT$ symmetry breaking terms are $\lambda_j\,r_j$ with $r_1=x\,,\,\,r_2=y$. The correspondent time-independent displaced Schroedinger equation for this potential is
\be \label{eq3.2}
-\frac{\hbar^2}{2m}\,{\cal D}_x^2\,\Psi-\frac{\hbar^2}{2m}\,{\cal D}_y^2\,\Psi+V(x,y)\Psi=E\Psi\,,
\ee
where 
\be \label{eq3.3}
{\cal D}_x=(\xi_1+\gamma_1\,x)\frac{\partial}{\partial\,x}+\beta_1\,\gamma_1\,;\qquad {\cal D}_y=(\xi_2+\gamma_2\,y)\frac{\partial}{\partial\,y}+\beta_2\,\gamma_2\,.
\ee
Note that here we are working with two generalized versions for ${\cal D}$ operator, once our model is non-Hermitian by construction we do not need to restrict ourself to Hermitian displacement momentum operators. Now, let us consider the following change of variables:
\be \label{eq3.4}
u_1=\xi_1+\gamma_1\,x\,; \qquad u_2=\xi_2+\gamma_2\,y\,,
\ee
yielding to the momentum operators
\be \label{eq3.5}
{\cal D}_x^2=u_1^2\,\gamma_1^2\,\frac{\partial^2}{\partial\,u_1^2}+a_1\,\gamma_1^2u_1\,\frac{\partial}{\partial\,u_1}+\beta_1^2\,\gamma_1^2\,;\qquad a_1=1+2\,\beta_1\,;
\ee
\be \label{eq3.6}
{\cal D}_y^2=u_2^2\,\gamma_2^2\,\frac{\partial^2}{\partial\,u_2^2}+a_2\,\gamma_2^2u_2\,\frac{\partial}{\partial\,u_2}+\beta_2^2\,\gamma_2^2\,;\qquad a_2=1+2\,\beta_2\,.
\ee
As a next step, let us work with the change of variables $u_j=\bar{z}_j+\xi_{j}$, with $j=1,2$. After some algebra the resultant differential equation is such that
\ben \label{eq3.7}
&&
(\bar{z}_1+\xi_{1})^2\gamma_1^2\,\frac{\partial^2\,\Psi}{\partial\,\bar{z}_1^2}+a_1\gamma_1^2(\bar{z}_1+\xi_{1})\frac{\partial\,\Psi}{\partial\,\bar{z}_1}+(\bar{z}_2+\xi_{2})^2\gamma_2^2\,\frac{\partial^2\,\Psi}{\partial\,\bar{z}_2^2}\\ \nonumber 
&&
+a_2\gamma_2^2(\bar{z}_2+\xi_{2})\frac{\partial\,\Psi}{\partial\,\bar{z}_2}-\bar{\omega}_1^2\,\bar{z}_1^2\Psi-\bar{\omega}_2^2\,\bar{z}_2^2\Psi -\frac{2m}{\hbar^2\,\gamma_1}(\lambda_1+i\,\sigma_1)\,\bar{z}_1\,\Psi\\ \nonumber
&&
-\frac{2m}{\hbar^2\,\gamma_2}(\lambda_2+i\,\sigma_2)\,\bar{z}_2\,\Psi +\beta_1^2\,\gamma_1^2\Psi+\beta_2^2\,\gamma_2^2\Psi=-\frac{2\,m}{\hbar^2}\,E\,\Psi\,,
\een
where
$
\bar{\omega}_j^2=\frac{m^2\,\omega_j^2 }{\hbar^2\,\gamma_j^2}\,.
$
Performing one more change of variables given by $\bar{z}_{j}=z_{j}+c_{j}$, where $c_{j}$ are complex constants, we find
\ben \label{eq3.8}
\nonumber
&&
\hspace{-0.5cm}(z_1+\xi_{1}+c_{1})^2\gamma_1^2\,\frac{\partial^2\,\Psi}{\partial\,z_1^2}+a_1\gamma_1^2(z_1+\xi_{1}+c_{1})\frac{\partial\,\Psi}{\partial\,z_1}+(z_2+\xi_2+c_2)^2\gamma_2^2\,\frac{\partial^2\,\Psi}{\partial\,z_2^2}\\ \nonumber 
&&
\hspace{-0.5cm}+a_2\gamma_2^2(z_2+\xi_2+c_2)\frac{\partial\,\Psi}{\partial\,z_2}-\bar{\omega}_1^2\,z_1^2\Psi-\bar{\omega}_2^2\,z_2^2\Psi-\bar{\omega}_1^2\,(2\,z_1\,c_1+c_1^2)\,\Psi\\ \nonumber
&&
\hspace{-0.5cm}-\bar{\omega}_2^2\,(2\,z_2\,c_2+c_2^2)\,\Psi-\frac{2m}{\hbar^2\,\gamma_1}(\lambda_1+i\,\sigma_1)\,(z_1+c_1)\,\Psi-\frac{2m}{\hbar^2\,\gamma_2}(\lambda_2+i\,\sigma_2) \\ 
&&
\hspace{-0.5cm}\times \,(z_2+c_2)\,\Psi+\beta_1^2\,\gamma_1^2\Psi+\beta_2^2\,\gamma_2^2\Psi=-\frac{2\,m}{\hbar^2}\,E\,\Psi\,.
\een
In order to eliminate the terms $z_1\,\Psi$, as well as,  $z_2\,\Psi$ we find the constraint
\be \label{eq3.10}
c_j=-\frac{\lambda_j+i\,\sigma_j}{m\,\omega_j^{\,2}}\,\gamma_j\,; \qquad j=1,2\,.
\ee
Besides, by defining  $\xi_{j}=1-c_{j}$, we yield to the differential equation
\ben \label{eq3.11}
&& \nonumber
(z_1+1)^2\gamma_1^2\,\frac{\partial^2\,\Psi}{\partial\,z_1^2}+a_1\gamma_1^2(z_1+1)\frac{\partial\,\Psi}{\partial\,z_1}+(z_2+1)^2\gamma_2^2\,\frac{\partial^2\,\Psi}{\partial\,z_2^2}+a_2\gamma_2^2(z_2+1)\frac{\partial\,\Psi}{\partial\,z_2}\\ \nonumber 
&&
-\bar{\omega}_1^2\,z_1^2\Psi-\bar{\omega}_2^2\,z_2^2\Psi=\bigg[-\frac{2\,m}{\hbar^2}\,E\,+\bar{\omega}_1^2\,c_1^2+\bar{\omega}_2^2\,c_2^2+\frac{2m}{\hbar^2\,\gamma_1}(\lambda_1+i\,\sigma_1)\,c_1 \\ 
&&
+\frac{2m}{\hbar^2\,\gamma_2}(\lambda_2+i\,\sigma_2)\,c_2-\beta_1^2\,\gamma_1^2-\beta_2^2\,\gamma_2^2\bigg]\Psi\,,
\een
where a real spectrum condition can be derived if we eliminate the imaginary terms of the right-hand side. Therefore, the real spectrum condition is such that
\ben \label{eq3.12}
\frac{\lambda_1\sigma_1}{\omega_1^{\,2}}=-\frac{\lambda_2\sigma_2}{\omega_2^{\,2}}\,.
\een
Then, our second order differential equation is written as
\ben \label{eq3.13}
&&
(z_1+1)^2\gamma_1^2\,\frac{\partial^2\,\Psi}{\partial\,z_1^2}+a_1\gamma_1^2(z_1+1)\frac{\partial\,\Psi}{\partial\,z_1}+(z_2+1)^2\gamma_2^2\,\frac{\partial^2\,\Psi}{\partial\,z_2^2} \\ \nonumber
&&
+a_2\gamma_2^2(z_2+1)\frac{\partial\,\Psi}{\partial\,z_2}-\bar{\omega}_1^2\,z_1^2\Psi-\bar{\omega}_2^2\,z_2^2\Psi
=\bar{E}\,\Psi\,,
\een
with
\be \label{eq3.14}
\bar{E}=-\frac{2\,m}{\hbar^2}\,E\,-\frac{\lambda_1^{\,2}-\sigma_1^{\,2}}{\hbar^2\,\omega_1^2}-\frac{\lambda_2^{\,2}-\sigma_2^{\,2}}{\hbar^2\,\omega_2^2}-\beta_1^2\,\gamma_1^2-\beta_2^2\,\gamma_2^2\,.
\ee
Let us now make a separation of variables for  $\Psi(z_1,z_2)$, which means to represent this wave function as $\Psi=\phi_1(z_1)\,\phi_2(z_2)$, besides the definition $\bar{E}=\bar{E}_1+\bar{E}_2$. Such an approach together with 
\be \label{eq3.15}
\phi_j(z_j)=\frac{\phi_j(z_j)}{(z_j+1)^{\,a_j/2}}\,;\qquad j=1,2\,
\ee
results in
\be \label{eq3.16}
(z_j+1)^2\,\phi_{j\,\,z_j\,z_j}-\widetilde{\omega}_j^2z_j^2\phi_j=\widetilde{E_j}\,\phi_j;\qquad j=1,2\,.
\ee
for
\be \label{eq3.17}
\widetilde{E_j}=\frac{\bar{E_j}}{\gamma_j^2}\,;\qquad \widetilde{\omega}_j^2=\frac{\bar{\omega}_j^2}{\gamma_j^2}\,.
\ee
As in the previous section, the solutions for both differential equations are given in terms of the associated Laguerre polynomials, whose explicit forms are
\be \label{eq3.18}
\phi_j=e^{-\widetilde{\omega_j}(z_j+1)}(z_j+1)^{\widetilde{\omega}_j-n_j}L_{n_j}^{\kappa_j}[2\widetilde{\omega}(z_j+1)]\,,
\ee
where $n_j$, and $\kappa_j$ are 
\be \label{eq3.19}
n_j=-\frac{1}{2} +\widetilde{\omega}_j-\frac{1}{2}\sqrt{1+4\widetilde{E_j}+4\widetilde{\omega_j}^{2}}\,;\qquad \kappa_j=\sqrt{1+4\widetilde{E_j}+4\widetilde{\omega_j}^{2}}\,.
\ee

Consequently, the general solution of this harmonic oscillator is
\ben \label{eq3.20}
&& \nonumber
\Psi(x,y)=e^{-\frac{m\omega_1}{\hbar\,\gamma_1^2}(\xi_1+\gamma_1\,x)}(\xi_1+\gamma_1\,x)^{\frac{m\,\omega_1}{\hbar\,\gamma_1^2}-n_1-\frac{1}{2}-\beta_1}L_{n_1}^{\kappa_1}\,\left[\frac{2\,m\,\omega_1}{\hbar\,\gamma_1^2}(\xi_1+\gamma_1\,x)\right]\\ 
&&
\times\, e^{-\frac{m\omega_2}{\hbar\,\gamma_2^2}(\xi_2+\gamma_2\,y)}(\xi_2+\gamma_2\,y)^{\frac{m\,\omega_2}{\hbar\,\gamma_2^2}-n_2-\frac{1}{2}-\beta_2}L_{n_2}^{\kappa_2}\,\left[\frac{2\,m\,\omega_2}{\hbar\,\gamma_2^2}(\xi_2+\gamma_2\,y)\right]\,.
\een
Analogously with the simple oscillator case, the $n_j$'s must be integers, unveiling the real spectrum
\ben \label{eq3.21}
&&
\widetilde{E_j}=n_j(n_j+1)-\widetilde{\omega_j}(1+2n_j)\,;\\ \nonumber
&&
E=\sum_{j=1}^{2}\left\{\,\hbar\omega_j\left(n_j+\frac{1}{2}\right)-\frac{\hbar^{2}\gamma_j^{2}}{2\,m}\left(n_j\left(n_j+1\right)+\beta_j^2\right)-\frac{\lambda_j^2-\sigma_j^2}{2m\omega_j}\right\}\,.
\een
Thus, we were able to find real spectrum for a displaced harmonic oscillator which breaks the $PT$ symmetry.  Despite the fact that we have a real spectrum for this oscillator, its non-Hermitian features still present through the parameters $\lambda_j$ and $\sigma_j$. 

\section{Statistical mechanics}
\label{sec_5}

As a matter of increase the physical interpretations for the displaced harmonic oscillators, we are going to perform a statistical treatment of these systems. Firstly, let us describe carefully the thermodynamic quantities for the simple displaced harmonic oscillator. The ingredient which is the foundation of any statistical analysis is the partition function, whose form for a canonical ensamble is
\be \label{eq4.1}
Z=\sum_{n=0}^{\infty}\,e^{\,-\beta\,E_{\,n}}\,; \qquad \beta=\frac{1}{k_{B}\,T}\,,
\ee
where $k_{B}$  is the Boltzmann constant. The main problem in find $Z$ for the displaced oscillator is that its spectrum results in a finite partition function only if $\gamma \rightarrow 0$, see \eqref{eq2.16}. This same issue was studied by Strekalov \cite{strekalov}, when he was investigating numerical and analytical partition functions for Morse oscillators. These oscillators described the rotation and vibrational effects on polyatomic molecules. 

In his studies, Strekalov chose a proper cut-off for the partition function, based on the calculation of the dissociation energy of the molecules. In order to determine such an energy, we need to establish a maximum energy state for the spectrum, which can be derived from  
\be \label{eq4.2}
\frac{d\,E_{\,n}}{d\,n}\Bigg|_{N}=0\,.
\ee
Then, we are able to determine the maximum allowed state for the displaced oscillator by substituting \eqref{eq2.16} in \eqref{eq4.2}, such a procedure yields to
\be \label{eq4.3}
N=\frac{m\,\omega}{\hbar\,\gamma^{\,2}}-\frac{1}{2}\,.
\ee
Therefore, the dissociation energy for the displaced oscillator is 
\be \label{eq4.4}
E_{\,d}=E_{\,N}-E_{\,0}\,,
\ee
where $E_{\,0}$ is the ground state energy. 

Once $N\geq0$, then $(\ref{eq4.3})$ unveils that
\be
\gamma^2\leq2\,\frac{m\,\omega}{\hbar}\,,
\ee
meaning a physical restriction for the allowed values of $\gamma$. Another possible restriction appears by fixing $N$ for a given molecule, for instance,  an iodine molecule has $N=173$ \cite{strekalov}, therefore, 
\be
\gamma^{\,2}=\frac{\hbar}{m\,\omega}\left(173+\frac{1}{2}\right)\,,
\ee
representing a measurement of internal interactions of such a molecule \cite{strekalov}.

Moreover, repeating the steps introduced in \cite{strekalov}, we are able to rewrite the partition function from \eqref{eq4.1} as
\be \label{eq4.5}
Z=\sum_{n=0}^{N}\,e^{-\beta\,E_{\,n}}\,.
\ee
One can note that for $\gamma \rightarrow 0$, Eq. $(\ref{eq4.3})$ informs that $N\rightarrow \infty$, therefore 
\be
Z=\sum_{n=0}^{N}\,e^{-\beta\,E_{\,n}} \qquad \rightarrow \qquad Z=\sum_{n=0}^{\infty}\,e^{-\beta\,E_{\,n}}\,; \qquad E_n=\hbar\,\omega\left(n+\frac{1}{2}\right)\,,
\ee
recovering $(\ref{eq4.1})$, which is consistent with the spectrum behavior presented in $(\ref{eq2.16})$. Let us take back to Eq. $(\ref{eq4.5})$, there we can observe that it can be expressed as
\be \label{eq4.5_1}
Z=Z_{\,H}\,\sum_{k=0}^{\infty}\,\left(\frac{\hbar^{\,2}\gamma^{\,2}\beta}{2\,m}\right)^{\,k}\,\frac{I_ {\,2\,k}}{k!}\,,
\ee
where
\be \label{eq4.6}
I_ {\,k}=\frac{1}{Z_{\,H}}\,\sum_{n=0}^{N}\,\left(n+\frac{1}{2}\right)^{\,k}\,e^{\,-\left(n+\frac{1}{2}\right)\,\beta\,\hbar\,\omega}\,; \qquad Z_{\,H}=\sum_{n=0}^{N}e^{-\left(n+\frac{1}{2}\right)\,\beta\,\hbar\,\omega}\,.
\ee
Moreover, we can rewrite the series over $k$ in Eq. \eqref{eq4.5_1} using the so-called cumulant expansion, introduced by Kubo in \cite{kubo}. This expansion results in 
\be \label{eq4.7}
\sum_{k=0}^{\infty}\,\left(\frac{\hbar^{\,2}\gamma^{\,2}\beta}{2\,m}\right)^{\,k}\,\frac{I_ {\,2\,k}}{k!}=\exp\,\left\{\sum_{k=1}^{\infty}\,\left(\frac{\hbar^{\,2}\gamma^{\,2}\beta}{2\,m}\right)^{\,k}\frac{\mu_{k}}{k!}\right\}\,,
\ee
where the values of  $\mu_{k}$ up to third-order in  $\beta$ are
\be \label{eq4.8}
\mu_{\,1}=I_{\,2}\,; \qquad \mu_{\,2}=I_{\,4}-I_{\,2}^{\,2}\,; \qquad \mu_{\,3}=I_{\,6}-3\,I_{\,2}\,I_{\,4}+2\,I_{\,2}^{\,3}\,.
\ee

\begin{figure}[ht!]
\centering
\includegraphics[width=0.4 \columnwidth]{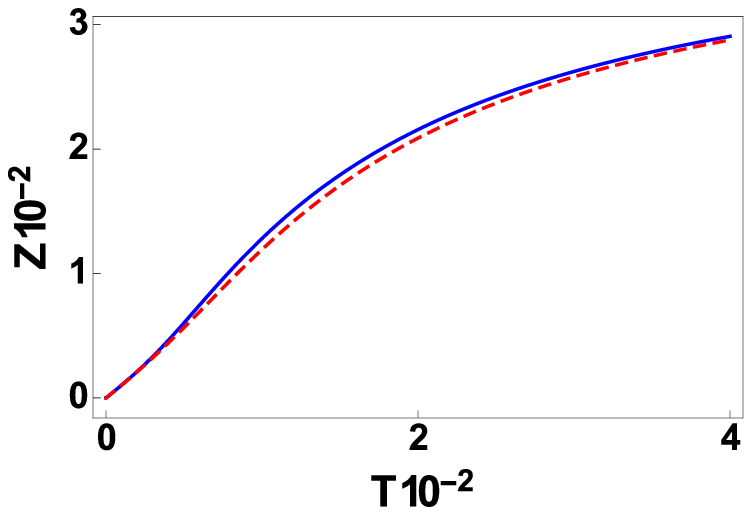} \hspace{0.5 cm} \includegraphics[width=0.4 \columnwidth]{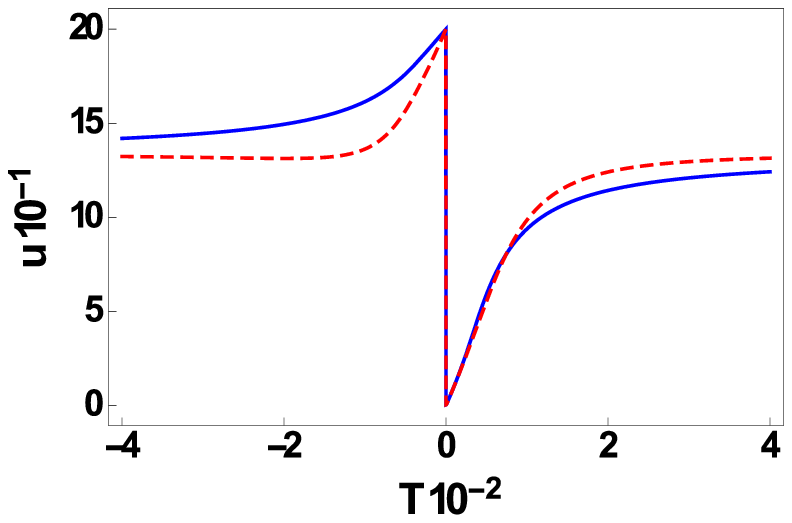}  
\caption{In the left panel we show the evolutions of the partition functions  $Z$ (solid blue curve), and $Z_{\,2}$ (dashed red curve) in respect to temperature. The right panel describes the behavior of the internal energies derived from $Z$ (solid blue curve), and from $Z_2$ (dashed red curve) when the temperature increases.}
\label{fig.1}
\end{figure}

\begin{figure}[h!]
\centering
\includegraphics[width=0.4 \columnwidth]{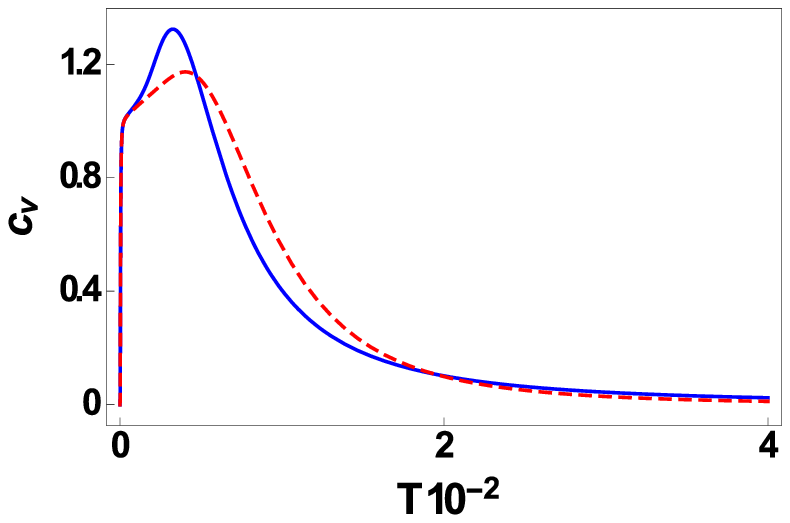} \hspace{0.2 cm}  \includegraphics[width=0.4 \columnwidth]{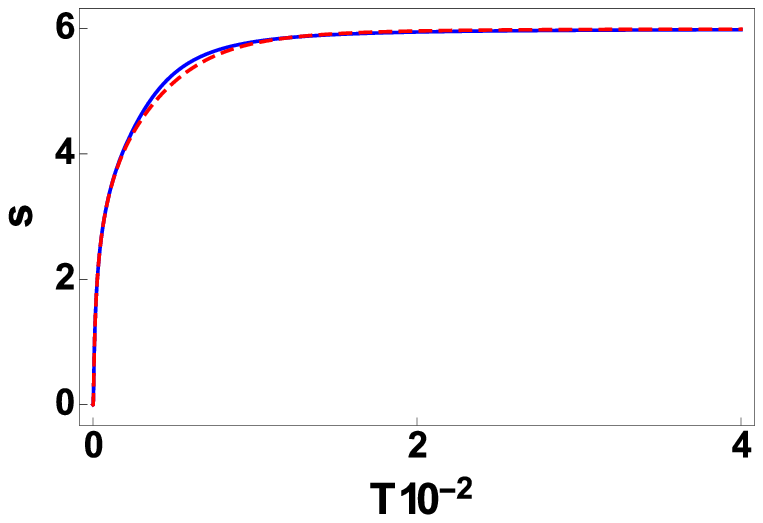}  
\caption{The left upper panel unveils the evolution of the specific heat related with  $Z$ (solid blue curve), and $Z_{\,2}$ (dashed red curve) in respect to temperature.  In the right panel we can observe the behavior of entropy  as function of the temperature for $Z$ (solid blue curve), and for  $Z_2$ (dashed red curve). }
\label{fig.2}
\end{figure}

\begin{figure}[ht!]
\centering
\includegraphics[width=0.4 \columnwidth]{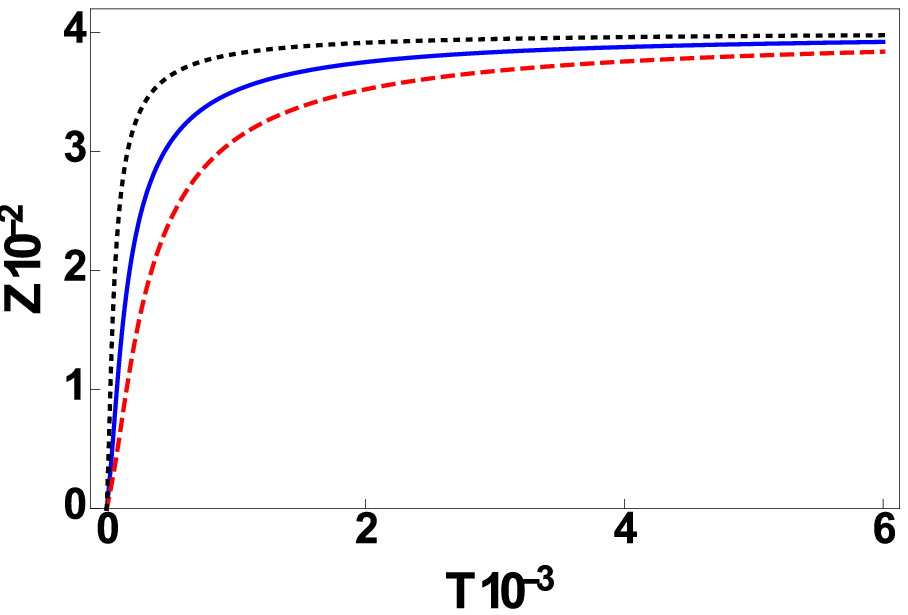} \hspace{0.5 cm} \includegraphics[width=0.4 \columnwidth]{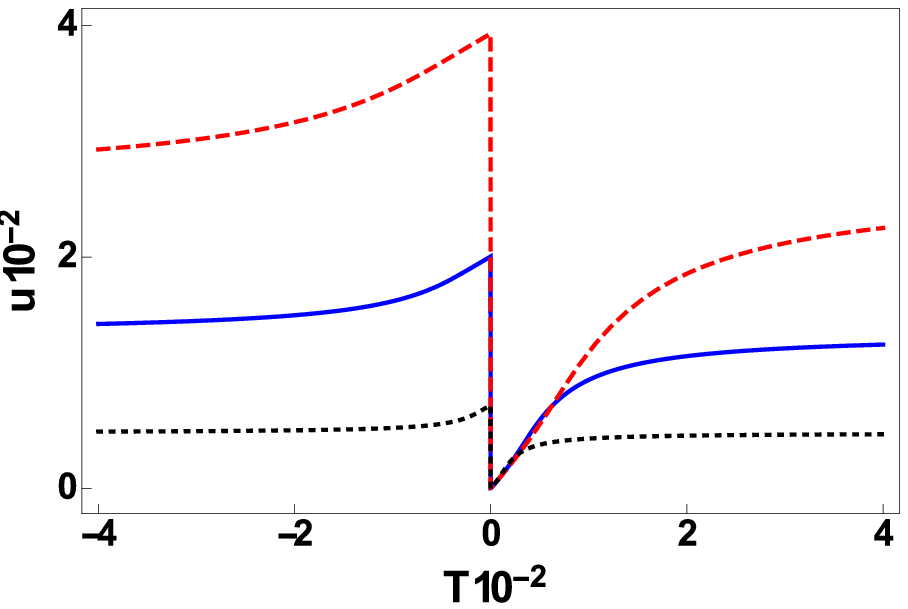}  
\caption{The left panel reveals the behavior of $Z$ for $\gamma=0.03$ (dotted black curve), $\gamma=0.05$ (solid blue curve), $\gamma=0.07$ (dashed red curve). Besides, in the right panel we show  the internal energy derived from $Z$ for $\gamma=0.03$ (dotted black curve), $\gamma=0.05$ (solid blue curve), $\gamma=0.07$ (dashed red curve). Both panels were plotted with $N=400$.}
\label{fig.1.1}
\end{figure}

\begin{figure}[h!]
\centering
\includegraphics[width=0.4 \columnwidth]{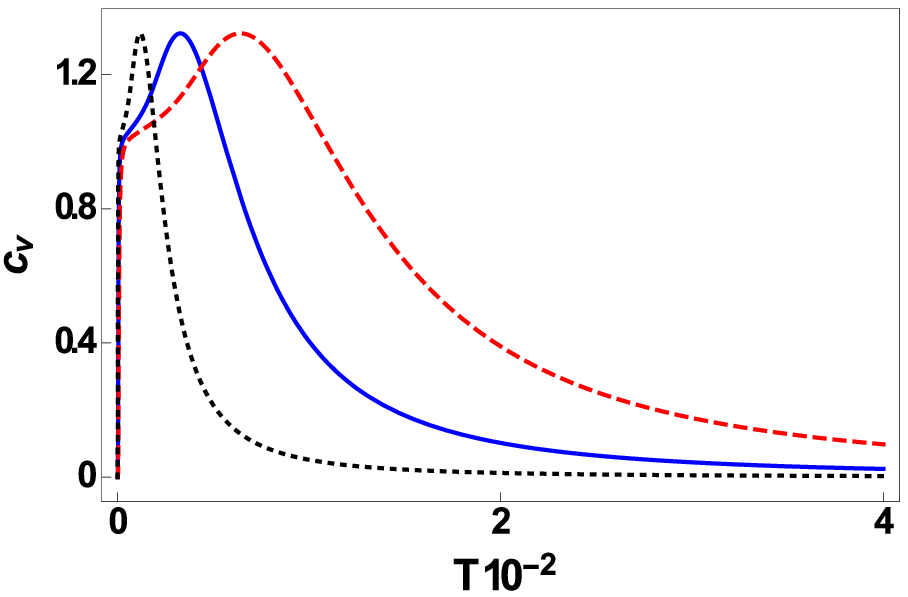} \hspace{0.2 cm}  \includegraphics[width=0.4 \columnwidth]{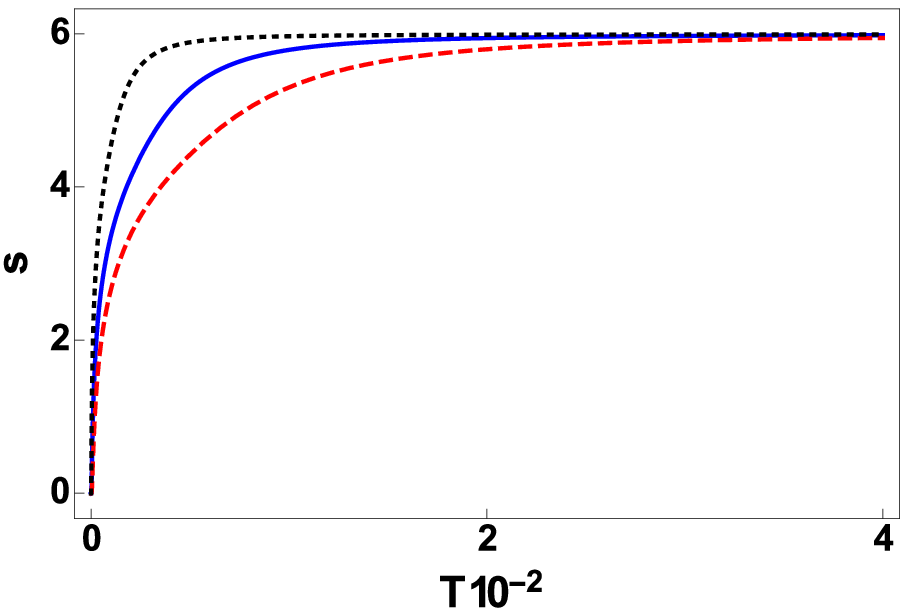}  
\caption{The left panel shows the features of the specific heat $c_v$ for $\gamma=0.03$ (dotted black curve), $\gamma=0.05$ (solid blue curve), $\gamma=0.07$ (dashed red curve). Besides, in the right panel we observe the entropy $s$ for $\gamma=0.03$ (dotted black curve), $\gamma=0.05$ (solid blue curve), $\gamma=0.07$ (dashed red curve). Both panels were plotted with $N=400$ and derived from the partition function $Z$. }
\label{fig.2.1}
\end{figure}

Then, the approximated version of the partition function is given by
\be \label{eq4.9}
Z_{2\,k} \approx Z_{\,H}\,\exp\,\left\{\left(\frac{\hbar^{\,2}\gamma^{\,2}\,\beta}{2\,m}\right)\,I_{\,2}+\left(\frac{\hbar^{\,2}\gamma^{\,2}\,\beta}{2\,m}\right)^{\,2}\,\frac{\left(I_{\,4}-I_{\,2}\right)^{\,2}}{2}+...\right\}\,.
\ee
The procedure above secures that the term $\gamma^{\,2}$ present in $E_{\,n}$ has a perturbative effect on the standard eigenvalues of the harmonic oscillator. Once we have the partition function in hands, we are able to use it together with the relations 
\be \label{eq4.10}
u=-\frac{\partial}{\partial\,\beta}\,\left(\log\,Z\right)\,; \,\,\, c_{v}=-k_{\,B}\beta^{\,2}\,\frac{\partial\,u}{\partial\,\beta}\,; \,\,\, s=k_{\,B}\,\log\,Z-k_{\,B}\,\beta\,\frac{\partial}{\partial\,\beta}\,\left(\log\,Z\right)\,,
\ee
to derive the internal energy, the specific heat, and the entropy, respectively.

Let us exemplify this approach by choosing  $\hbar=m=\omega=1$, and $\gamma=0.05$. Substituting such values in \eqref{eq4.3} yields to $N=399.5$, which means that we are going to work with a total of $N=400$ terms in the sum of the partition function. Taking the previous parameters into Eqs.  \eqref{eq4.5}, \eqref{eq4.5_1}, \eqref{eq4.6}, and \eqref{eq4.9}, results in
\be \label{eq4.11}
Z=\sum_{n=0}^{400}\,e^{\,-\beta\,E_{n}}\,; \qquad Z_{\,2}=Z_{\,H}\,\exp\,\left\{\left(\frac{\beta}{2}\right)\,I_{\,2}\right\}\,, 
\ee
whose forms are shown in the left panel of Fig. \ref{fig.1}, where we can see that both partitions functions become closer as the temperature increases.

The application of both partition functions from \eqref{eq4.11} in \eqref{eq4.10} yields to the graphics presented in Figs. \ref{fig.1}, and \ref{fig.2}. Such figures unveil that the thermodynamic quantities have the same features of two energy level systems, as depicted in the right-hand side of Fig. \ref{fig.1}. The left upper panel from Fig. \ref{fig.2} shows that the specific heat for $Z$  and for $Z_2$ display broad maximums, which is an anomalous behavior of the specific heat known as Schottky anomaly \cite{salinas}. This kind of anomaly is typical of systems with a limited number of energy levels  \cite{salinas}. It is also an interesting example of how microscopic quantum phenomena result in macroscopic effects. This anomaly appears in paramagnetic salts, in some ferromagnetic metals and also in minerals \cite{ventura,aronson}

The features of the entropy, plotted in the lower panel of Fig. \ref{fig.2}, corroborate with the Schottky anomaly observed in the graphics for the specific heat previously mentioned. There we can see that $s$ goes to zero for lower values of temperature. In such a regime,  the system is in its ground state configuration. Moreover, $s$  goes to a constant value for larger values of $T$. This constant value  means that the system has made its transition to the other allowed energy level and coincides with the region where the specific heat goes to zero.

As one can see, the entropy for both $Z$ and $Z_2$ increases from zero at an elevated rate, then, after reach the temperature related with the maximums of the specific heats, the entropies still growing but now with a lower rate. Moreover, we also note that the thermodynamic quantities derived from $Z$ and $Z_2$ are equivalent for higher values of temperature, however some quantities like the energy density, the specific heat, and the entropy are sensible to the cut-off procedure as the temperature falls down. 

In order to complete our discussions about the thermodynamic quantities, we built Figs. \ref{fig.1.1} and \ref{fig.2.1}. These sets of graphics reveal the partition function, the energy density, the specific heat and the entropy for different values of $\gamma$. The figures were generated with $N=400$, $m=1$, $\hbar=1$, and present different values for the frequency $\omega$, which were determined from $(\ref{eq4.3})$ as follows
\be
\omega=\left(N+\frac{1}{2}\right)\,\gamma^{\,2}\,; \qquad N=400\,.
\ee
In Fig \ref{fig.1.1} we can observe that the partition functions go to a constant as $T$ gets higher, and this transition occurs for smaller values of $T$ as lower values of $\gamma$ are taken. Moreover, the energy densities become higher for bigger values of $\gamma$, representing an elevation of the energy gap for the two level systems. Besides, the graphics of Fig. \ref{fig.2.1} corroborate with the behavior of $Z$ and $u$. There, we realize that the specific heats have abrupt transitions between the energy levels as $\gamma$ gets smaller. The same features can be appreciated for the entropy $s$.

In his work Strekalov treats the parameter related with the Morse potential as measurement of anharmonicity \cite{strekalov}, which is interpreted as an effect due intermolecular interactions. Then, if we trace an analogy between the q-algebra parameter with the Strekalov one, we can verify that the transition between the different energy levels is more abrupt for smaller anharmonic effects (or for lower internal interactions of the system).

An equivalent approach can be performed to the displaced anisotropic harmonic oscillator. Analogously with our previous procedures, the partition function for this oscillator can be derived from
\be \label{eq4.12}
Z=\sum_{n_1=0}^{N_1}\,\sum_{n_2=0}^{N_2}\,e^{-\beta\,\left(E_{n_1}+E_{n_2}\right)}\,,
\ee
where 
\ben \label{eq4.13}
&&
E_{n_j}=\,\hbar\omega_j\left(n_j+\frac{1}{2}\right)-\frac{\hbar^{2}\gamma_j^{2}}{2\,m}\left(n_j\left(n_j+1\right)+\beta_j^2\right)-\frac{\lambda_j^2-\sigma_j^2}{2m\omega_j}\,; \\ \nonumber
&&
 N_j= \frac{m\,\omega_j}{\hbar\,\gamma_{\,j}^{\,2}}-\frac{1}{2}\,,
\een
with $j=1,2$. By repeating the cut-off approach adopted by Strekalov in \cite{strekalov}, we are able to rewrite the partition function as
\ben \label{eq4.14}
&&
Z=Z_{\,H_1}\,Z_{\,H_2}\,\sum_{k_1=0}^{\infty}\,\sum_{k_2=0}^{\infty}\left(\frac{\hbar^{\,2}\gamma_1^{\,2}\,\beta}{2\,m}\right)^{\,k_1}\,\left(\frac{\hbar^{\,2}\gamma_2^{\,2}\,\beta}{2\,m}\right)^{\,k_2}\, \\ \nonumber
&&
\times\,\frac{I_{k_1}}{k_1!}\,\frac{I_{k_2}}{k_2!}\,\exp\left\{\frac{\hbar^{\,2}\beta}{2\,m}\,\left(\gamma_1^{\,2}\,\beta_{1}^{\,2}+\gamma_2^{\,2}\,\beta_{2}^{\,2}\right)+\frac{\beta}{2\,m}\,\left(\frac{\lambda_1^{\,2}-\sigma_1^{\,2}}{\omega_1}+\frac{\lambda_2^{\,2}-\sigma_2^{\,2}}{\omega_2}\right)\right\}\,,
\een
where
\be \label{eq4.15}
I_{k_j}=\frac{1}{Z_{\,H_j}}\,\sum_{n_j=0}^{N_j}\,\left(n_{j}\,(n_{j}+1)\right)^{\,k_j}\,e^{\,-\beta\,\left(n_{j}+\frac{1}{2}\right)\,\hbar\,\omega_{j}}\,; \,\,\, Z_{H_j}=\sum_{n_j=0}^{N}\,e^{-\beta\,\left(n_j+\frac{1}{2}\right)\,\hbar\,\omega_j}\,,
\ee
yielding to
\ben \label{eq4.16}
&&
Z_{\,1\,1}=Z_{\,H_1}\,Z_{\,H_2}\,\exp\,\left\{\frac{\hbar^{\,2}\,\gamma_1^{\,2}\,\beta}{2\,m}\,I_{1_1} + \frac{\hbar^{\,2}\,\gamma_2^{\,2}\,\beta}{2\,m}\,I_{1_2}\right\} \\ \nonumber
&&
\times\,\exp\left\{\frac{\hbar^{\,2}\beta}{2\,m}\,\left(\gamma_1^{\,2}\,\beta_{1}^{\,2}+\gamma_2^{\,2}\,\beta_{2}^{\,2}\right)+\frac{\beta}{2\,m}\,\left(\frac{\lambda_1^{\,2}-\sigma_1^{\,2}}{\omega_1}+\frac{\lambda_2^{\,2}-\sigma_2^{\,2}}{\omega_2}\right)\right\}\,,
\een
as a first-order approximation in $\beta$.

\begin{figure}[ht!]
\centering
\includegraphics[width=0.4 \columnwidth]{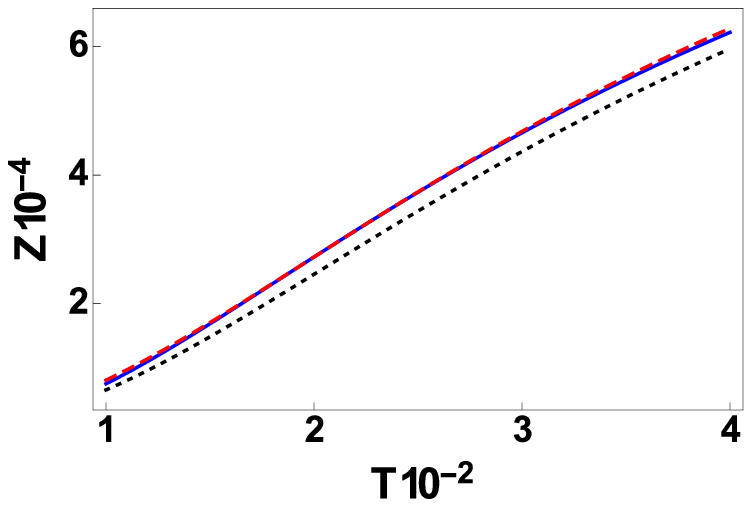} \hspace{0.5 cm} \includegraphics[width=0.4 \columnwidth]{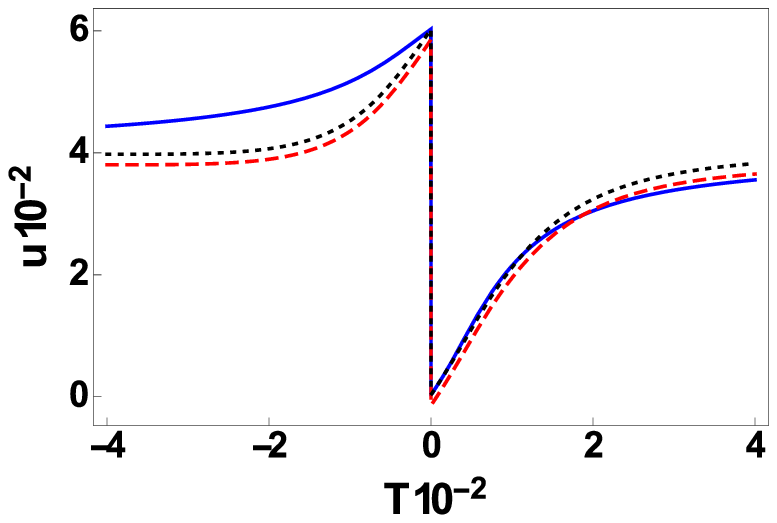}
\caption{The left panel describes the evolution of the partition functions $Z$ (solid blue curve), $Z_{\,1\,1}$ (dashed red curve),and $Z_{\,1\,1}$ (dotted black curve) in respect to the temperature. The right panel shows the internal energies as functions of temperature for $Z$ (solid blue curve), $Z_{\,1\,1}$ (dashed red curve), and $Z_{\,1\,1}$ (dotted black curve). The solid blue and the dotted black curves were depicted with  $\lambda_1=\lambda_2=0.5$, and $\sigma_1=\sigma_2=2$, besides the dashed red curves were plotted with $\lambda_1=\lambda_2=5$, and $\sigma_1=\sigma_2=1$.}
\label{fig.3}
\end{figure}

\begin{figure}[h!]
\centering
\includegraphics[width=0.4 \columnwidth]{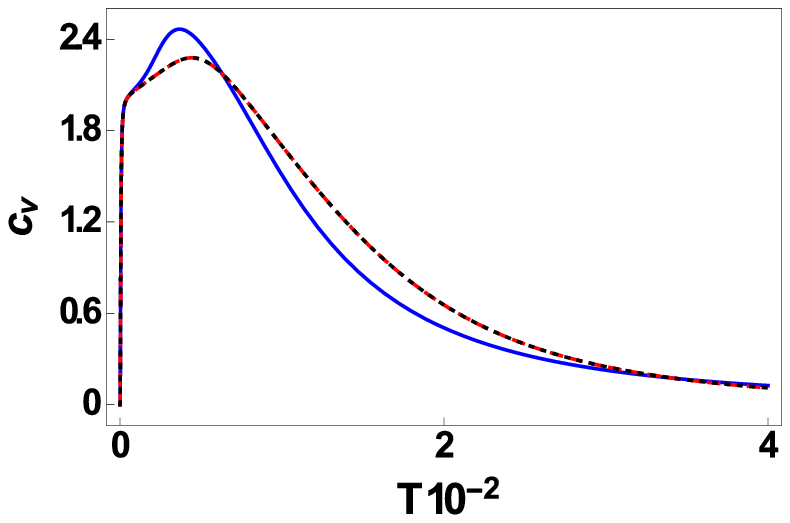} \hspace{0.2 cm}  \includegraphics[width=0.4 \columnwidth]{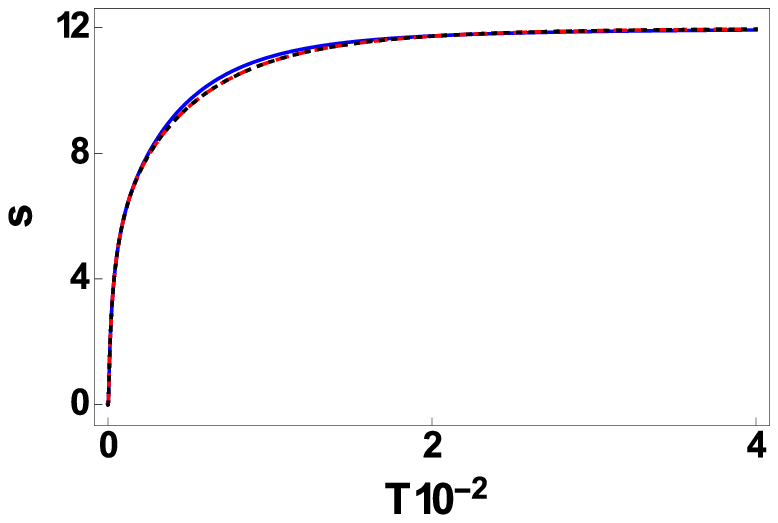}
\caption{The left upper panel unveils the evolution of the specific heat in respect to the temperature, derived from $Z$ (solid blue curve), $Z_{\,1\,1}$ (dashed red curve), and $Z_{\,1\,1}$ (dotted black curve). The right panel unveils the entropies from $Z$ (solid blue curve), $Z_{\,1\,1}$ (dashed red curve), and $Z_{\,1\,1}$ (dotted black curve) as the temperature increases.  The solid blue and the dotted black curves were depicted with  $\lambda_1=\lambda_2=0.5$, and $\sigma_1=\sigma_2=2$, besides the dashed red curves were plotted with $\lambda_1=\lambda_2=5$, and $\sigma_1=\sigma_2=1$.}
\label{fig.4}
\end{figure}

In order to exemplify the properties of the thermodynamic quantities of this oscillator, let us consider  $m=1$, $\hbar=1$, $\gamma_1=0.05$ $\gamma_2=0.05\,\sqrt{2}$, $\omega_1=1$, $\omega_2=2$, imposing $N_1=N_2=399.5$. Then, we adopted $N_1=N_2=400$ as the cut-off for our partition function. By substituting these ingredients into $Z$ \eqref{eq4.12}, and into $Z_{\,1\,1}$ \eqref{eq4.16}, we derive the partition functions depicted in the left panel of Fig.  \ref{fig.3}. Besides, these partition functions allowed us to derive the quantities exhibited in Figs. \ref{fig.3} e \ref{fig.4}.

\begin{figure}[ht!]
\centering
\includegraphics[width=0.4 \columnwidth]{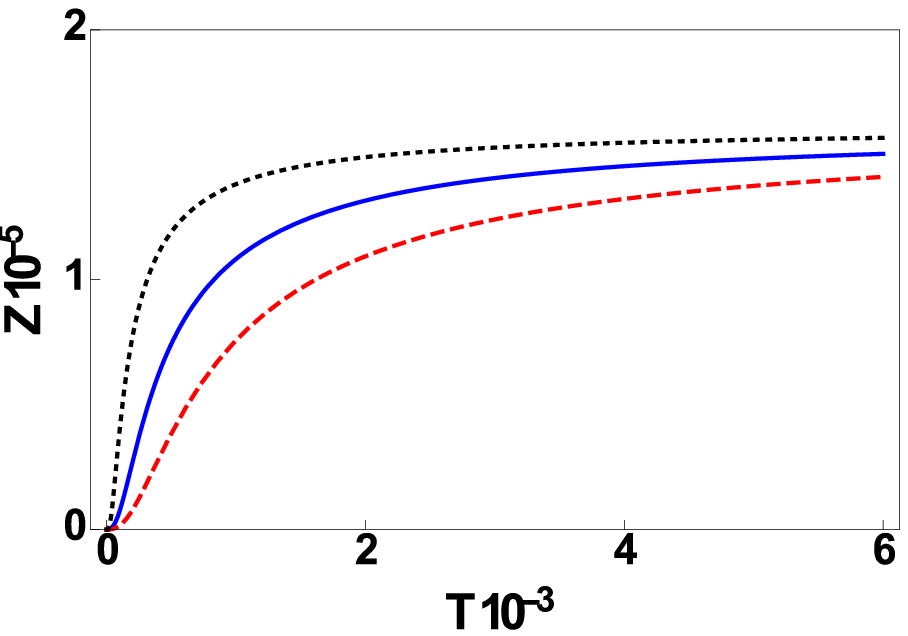} \hspace{0.5 cm} \includegraphics[width=0.4 \columnwidth]{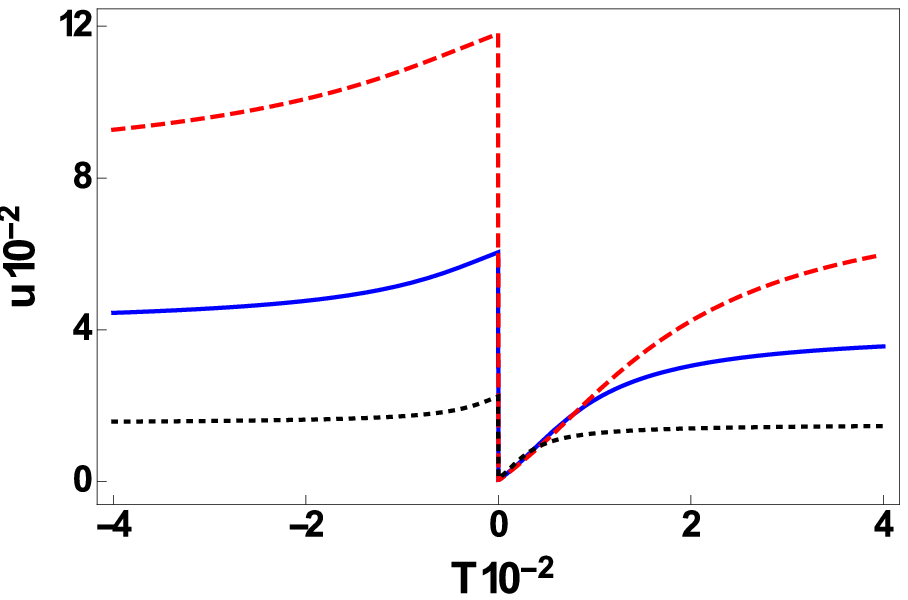}
\caption{The left panel reveals the behavior of $Z$ for $\gamma_1=0.03$ and $\gamma_2=0.03\,\sqrt{2}$ (dotted black curve), $\gamma_1=0.05$ and $\gamma_2=0.05\,\sqrt{2}$ (solid blue curve), $\gamma_1=0.07$ and $\gamma_2=0.07\,\sqrt{2}$ (dashed red curve). Besides, in the right panel we show  the internal energy derived from $Z$ for $\gamma_1=0.03$ and $\gamma_2=0.03\,\sqrt{2}$ (dotted black curve), $\gamma_1=0.05$ and $\gamma_2=0.05\,\sqrt{2}$ (solid blue curve), $\gamma_1=0.07$ and $\gamma_2=0.07\,\sqrt{2}$ (dashed red curve). Both panels were plotted with $N_1=N_2=400$, $\lambda_1=\lambda_2=0.5$, and $\sigma_1=\sigma_2=2$.}
\label{fig.3.1}
\end{figure}

\begin{figure}[h!]
\centering
\includegraphics[width=0.4 \columnwidth]{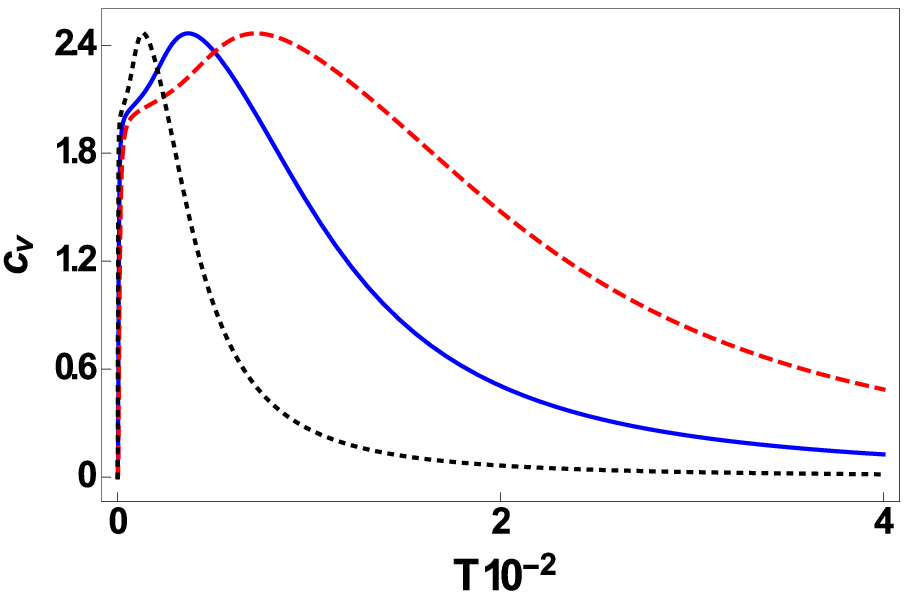} \hspace{0.2 cm}  \includegraphics[width=0.4 \columnwidth]{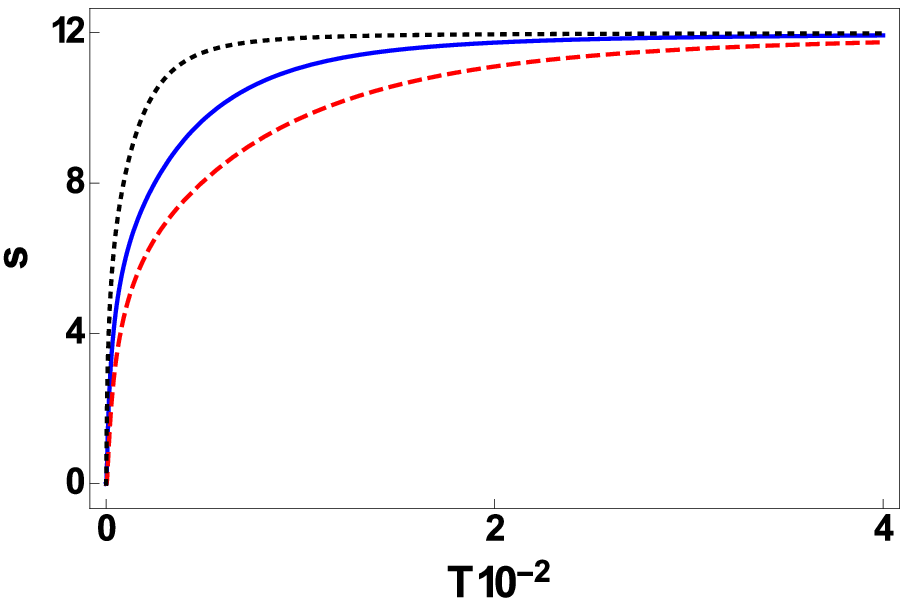}
\caption{The left panel shows the features of the specific heat $c_v$ for $\gamma_1=0.03$ and $\gamma_2=0.03\,\sqrt{2}$ (dotted black curve), $\gamma_1=0.05$ and $\gamma_2=0.05\,\sqrt{2}$ (solid blue curve), $\gamma_1=0.07$ and $\gamma_2=0.07\,\sqrt{2}$ (dashed red curve). Besides, in the right panel we observe the entropy $s$ with $\gamma_1=0.03$ and $\gamma_2=0.03\,\sqrt{2}$ (dotted black curve), $\gamma_1=0.05$ and $\gamma_2=0.05\,\sqrt{2}$ (solid blue curve), $\gamma_1=0.07$ and $\gamma_2=0.07\,\sqrt{2}$ (dashed red curve). Both panels were plotted with $N_1=N_2=400$   $\lambda_1=\lambda_2=0.5$, $\sigma_1=\sigma_2=2$, and derived from the partition function $Z$.}
\label{fig.4.1}
\end{figure}

There, the solid blue and dotted black curves were plotted with $\beta_1=\beta_2=0.5$, $\lambda_1=\lambda_2=0.5$, $\sigma_1=\sigma_2=2$, and from $Z$, and $Z_{\,1\,1}$, respectively. Moreover, the dotted red curves were depicted from $Z_{\,1\,1}$ with $\beta_1=\beta_2=0.5$, $\lambda_1=\lambda_2=5$, and $\sigma_1=\sigma_2=1$. We can note that the real spectral condition lead us to thermodynamic quantities for this displaced non-hermitian oscillator, which are compatible with the ones shown in Figs. \ref{fig.1} and \ref{fig.2}. 
 One more time, in order to complete the description of the thermodynamic quantities we plotted Figs. \ref{fig.3.1} and \ref{fig.4.1}. The graphics were depicted for partition function $Z$, keeping $N_1=N_2=400$ and changing the values of $\gamma_1$ and $\gamma_2$. We are able to conclude that the evolution of the thermodynamic quantities in respect to temperature $T$ corroborates with the behaviors presented in Figs. \ref{fig.1.1} and \ref{fig.2.1}.

Furthermore, the $PT$ symmetry breaking parameters $\lambda_j$ and $\sigma_j$, allows a fine tunning adjustment for the partition functions as well as for the thermodynamic quantities, despite the specific heat, which stills sensible to the cut-off procedure. Such a fine tunning property is a remarkable tool to adjust the displaced oscillator in the description of experimental data.

\section{Final remarks}
\label{sec_6}

In this study we developed a new route to justify the Hermitian version of the displacement operator introduced in \cite{r1,r2}. We shown that the Weyl ordering is crucial in order to derive a Hermitian version of this operator, complementing the work from Mazharimousavi \cite{r2}. We also introduced a new type of displaced oscillator which breaks both $PT$ symmetry and  hermiticity. The analytical energy spectrum and the waves functions were derived, besides the so-called real spectrum condition. This model shows an interesting example of generalization of the non-Hermitian model introduced by Bender and Boettcher in its seminal paper \cite{bender}. Moreover we were able to derive several thermodynamic quantities for a simple displaced harmonic oscillator as well as for the displaced anisotropic two-dimensional non-Hermitian harmonic oscillator. The thermodynamic quantities here found, corroborate with the statistical description of a two-level system, and the parameter $\gamma$ can be viewed as internal interactions of the system, following the interpretation of \cite{strekalov}. 

The graphics for the specific heat and for the entropy of both oscillators can be compared with several experimental data related with the Schottky anomaly, we can mention for instance, the investigations of magnetic excitations in antiferromagnetic and paramagnetic phases of polycrystalline $Fe_2SiO_4$ \cite{aronson}. Other example of application of such a procedure is in theoretical description of the iridium double perovskite $Sr_2YIrO_6$ \cite{corredor}, or for the thermodynamic analysis of the so-called spin ice materials such as $Dy_2Ti_2O_7$ \cite{bramwell}. Besides, another potential application of these models is in the thermodynamic features of the paramagnetic salt Cerium-Magnesium-Nitride as recently studied in \cite{sereni}.

The methodology introduced can be implemented in other quantum Hamiltonian operators, such as quantum phases in dipole particles \cite{wei}, or in quantum states in electromagnetic fields for rotating space-time \cite{konno}. Another interesting generalization would be based on the application of our procedure to generate systems with  multiple Schottky anomalies. These multiple anomalies appear in several experimental data, like in the  heavy-fermion compound $Ce_3Pd_{20}Si_6$ \cite{mariano} or for the intermetallic compound $YbPt_2Sn$ \cite{jang}.  We hope to report on such contributions in near future.

\section*{Acknowledgements} 

The authors would like to thank CNPq, and CAPES (Brazilian agencies) for support, and also the anonymous reviewers for their guidance and suggestions during the submission process.

\end{document}